\documentclass[conference]{IEEEtran}
\IEEEoverridecommandlockouts
\usepackage{graphicx}
\graphicspath{{figures/}}
\usepackage[caption=false,font=footnotesize]{subfig}
\usepackage{float}

\usepackage{cite}

\usepackage{amsmath,amssymb,amsfonts}
\usepackage{textcomp}
\usepackage{xcolor}
\usepackage{romannum}
\usepackage[linesnumbered,ruled,vlined]{algorithm2e}

\makeatletter
\newcommand{\removelatexerror}{\let\@latex@error\@gobble}
\makeatother
\usepackage{algpseudocode}

\SetKwInput{KwInput}{START}                
\SetKwInput{KwOutput}{Output}              

\def\BibTeX{{\rm B\kern-.05em{\sc i\kern-.025em b}\kern-.08em
    T\kern-.1667em\lower.7ex\hbox{E}\kern-.125emX}}
\begin{document}

\title{A Spectrum Aware Mobility Pattern Based Routing Protocol for CR-VANETs}

\author{\IEEEauthorblockN{Sharmin Akter$^{1}$ and Nafees Mansoor$^{2}$}
\IEEEauthorblockA{Department of Computer Science and Engineering, University of Liberal Arts Bangladesh}
Email:{sharmin.akter2.cse@ulab.edu.bd$^{1}$, nafees@ieee.org$^{2}$}
}

\maketitle

\begin{abstract}
Cognitive radio technology offers an important function in the efficient utilization of the radio spectrum. Besides, it is expected that CR-enabled vehicular ad-hoc networks (CR-VANETs) enrich the communication performance of the existing vehicular networks (VANETs). However, to ensure efficient performance in a multi-hop communication, the routing protocol in CR-VANETs needs to consider the autonomous mobility of the vehicles as well as the stochastic availability of the channels. Hence, this paper introduces a spectrum-aware mobility pattern based reactive routing protocol for CR-VANET. The proposed protocol accommodates the dynamic behavior and selects a stable transmission path from a source node to the destination. Therefore, the proposed protocol is outlined as a weighted graph problem where the weight for an edge is measured based on a parameter termed as NHDF (Next-hop Determination Factor). The NHDF implicitly considers mobility patterns of the nodes and channel availability to select the optimum path for communication. Therefore, in the proposed routing protocol, the mobility pattern of a node is defined from the viewpoint of distance, speed, direction, and node's reliability. Furthermore, the spectrum awareness in the proposed protocol is measured over the number of shared common channels and the channel quality. It is anticipated that the proposed protocol shows efficient routing performance by selecting stable and secured paths from source to destination. Simulation is carried out to assess the performance of the protocol where it is witnessed that the proposed routing protocol outperforms existing ones.

\end{abstract}


\section{Introduction}
With the speedy advancement of wireless technologies, endless applications are expected to be deployed in every sphere of modern life. However, the radio spectrum scarcity becomes one of the major concerns with the amplifying demand in wireless applications. In such a situation, cognitive radio technology plays a major part by utilizing the radio spectrum efficiently \cite{Mitola}. Coined by J. Mitola III, Cognitive Radio Network (CRN)  is an intelligent wireless system that gathers information of the radio environment and controls its operational parameters i.e., transmission power, modulation strategy, etc. accordingly \cite{Nafees_Survey}. Primary User (PU) and Secondary User (SU) are two types of users involved in the CR system, where the PUs are the licensed users and SUs are the unlicensed users. Hence, in CRN, SU has to use the spectrum opportunistically without interrupting the PU's communication. 

On the other hand, the growing number of wheels on the road has outlined the necessity for the Intelligent Transportation System (ITS) to avoid the collision, monitor traffic, ensure road safety, etc. Consequently, vehicular ad-hoc network (VANET) has appeared to be the technology for these emerging vehicular applications by ensuring Vehicle to Vehicle (V2V) communication. In, the vehicular nodes are equipped with on-board units for sensing, transmitting and receiving messages \cite{Sharef}. This on-board arrangement in the vehicle provides appropriate decisions to avoid any disturbance or mishaps based on various information such as traffic status, vehicles' speed, etc. Moreover, it is anticipated that the CR-enabled vehicular network shows enriched communication efficiency than existing vehicular networks (VANETs).

Besides, in CR-VANET, the performance of the network highly depends on the underlying routing protocols. Moreover, similar to any other CRN, one of the important aspects of routing in CR-VANET is to utilize spectrum efficiently for vehicular communication. Furthermore, due to the dynamic vehicular movements and the changing availability of the radio channels, routing in CR-VANETs is quite challenging. Consequently, communication researchers are showing profound interest over the last few years on the routing protocols for CR-VANETs. Hence, due to the dynamic altering topology, a reliable and efficient route selection for multi-hop communication still remains an open research issue.

This paper presents a routing protocol for CR-VANET that is defined as a weighted graph problem. Distance, speed, direction, link quality, and reliability are considered as parameters for route selection in the proposed reactive protocol to enable faster and reliable data communication. The proposed routing protocol introduces a parameter called $NHDF$ (Next-hop Determination Factor) to select a routing path. Here, $NHDF$ value is associated with the number of common idle channels and transmitting weight value ($\xi_T$). Hence, by calculating distance through $RSSI$ (Received Signal Strength Indicator), the source node can estimate the location of the neighbor closest to forwarding the data to the destination. Vehicular speed and direction are measured for improving the performance during high mobility as the location of the nodes may vary, so this may cause link failure and packet drop. In this case, by calculating speed and angle, the neighboring node’s location can be predicted in the future and the packet is transmitted to the next neighbor. In the proposed routing protocol, link quality has been defined from the viewpoint of delay where three types of delays are proposed namely back-off, switching and queuing delay for estimating link weight which ensures route stability with a minimum link failure. A Reliable route selection mechanism is also proposed for detecting a selfish node or any external threads so that the neighboring nodes can trust received messages during the exchange of data. Route maintenance is also described in the proposed routing protocol when multiple routes are discovered during route selection or any link failure occurs during data transmission. To evaluate the performance of the proposed protocol, simulation results are compared with other protocols in terms of throughput, packet delivery ratio, and communication delay.

The rest of the article is organized as follows. Section \Romannum{2} depicts a brief discussion on existing routing protocols in CR-VANET and highlights their strengths and limitations based on the routing metrics. A network model for the proposed protocol is discussed in section \Romannum{3} and Section \Romannum{4} explains the proposed routing protocol. Next, simulation results and performance analyses of the proposed protocol are discussed in section \Romannum{5}. Finally, the paper ends with the conclusion and future work in section \Romannum{6}.

\section{RELATED WORK}
Noteworthy research works have been conducted over the past few years and different routing protocols are proposed to meet the challenges in cognitive radio enabled vehicular ad hoc networks. In this section, some of these recently developed routing protocols along with their limitations are discussed.

Using a trusted routing protocol algorithm technique, an optimized node selection Routing Protocol (ONSRP) is proposed in \cite{Jeyaprakash} for secured communication considering the distance, velocity, direction, and trust value. This protocol attempts to decrease link loss and overcomes the end-to-end delay issue. However, the ONSRP ignores high traffic density which causes channel congestion and overlooks the route maintenance process. Considering the greedy predictive forwarding strategy and perimeter forwarding strategy, a routing protocol is presented in \cite{Karimi} that predicts the neighbor vehicles’ potential position. However, link quality and external threads are not considered in this protocol for measuring any delay during communication and not able to predict future link failures during communication. Based on the node’s location, density and digital map Software-Defined Networking (SDN) based routing has been presented in \cite{Ji}. The high-density parameter has been used to solve the congestion problem. However, security issues have not been proposed in this protocol. Also, the prediction of link stability between two vehicular neighbors has been ignored. Based on Vehicular FoG Computing (VFC) architecture, a position based routing protocol is proposed in \cite{Ullah} initiating a dynamic route. This protocol can estimate the location of the neighbor nodes and also predict their position in the future for high mobility. Nevertheless, it lacks estimating link quality between nodes. Also, the authentication process and no recovery route mechanism for link failure are proposed.

Furthermore, the protocol presented in \cite{Ghaffari} can estimate link failure by evaluating link quality and predict the location of the nodes using density, mobility, and speed. However, any external attacks or threads may cause packet drop, data duplication or communication jam as this protocol fails to ensure secure and reliable communication. Intelligent based forwarding routing protocol proposed in \cite{Chahal} considers the parameters such as distance, moving direction, velocity and also link during to estimate the link quality ensuring stable communication without packet drop. But traffic density, high variable speed and security challenges regarding attacks or threats have been ignored. The protocol presented in \cite{Qureshi} considers road lengths, vehicle velocities, and distance for improving prediction of the future position of a neighboring vehicle. It limits the performance during low-density traffic. It also ignores the changing direction of the neighbor nodes and selfish node behavior.

Combining both predictive and perimeter forwarding approaches, the protocol in  \cite{Ye} can predict neighboring vehicles' position by analyzing acceleration, velocity, moving direction, distance and driver's intentions which achieves better performance eliminating local maximum problem improving overall performance. But traffic density, link quality between nodes and external attacks or threads such parameters have been ignored in this proposed routing protocol.  

From the literature, it is witnessed that existing protocols have resolved several major challenging aspects for routing in CR-VANET. However, a robust protocol considering the dynamic mobility patterns of the vehicles and varying availability of the channels is still absent\cite{Singh} for reliable communication. Therefore, addressing the current limitations while ensuring improved performance, this paper introduces a spectrum aware mobility-pattern based routing protocol in CR-VANET. Hence, the following section discusses the network model for the development of the proposed protocol.

\section{Network Model}
The proposed routing protocol is a multi-hop reactive protocol for CR-VANET where the intermediate nodes between the source and the destination are considered as the relay nodes. In the network, each vehicular node is equipped with the On-Board Unit (OBU) along with the omnidirectional antenna of a similar radio transmission range. It is assumed that each node holds the computational capability to calculate its $NHDF$ and aware of the neighbors' $NHDFs$. Next, this protocol enables the source node to select a path with a maximum cumulative of $NHDF$ to be the route for communication. In the network, any vehicular node senses and utilizes available channels opportunistically over the non-overlapping orthogonal channel set. Vehicular nodes in the network recognize the PU's transmission by observing the adjacent radio environment. Hence, it is assumed that the nodes use the energy detector based spectrum sensing approach to recognize the spectrum availability. Furthermore, the spatial variance and stochastic nature of the spectrum diversify the available channel sets of the vehicular nodes. 

Moreover, each node estimates the location of the neighbors based on the RSSI value. Here, each node waits for a random back-off time to reply on a $RREQ$ request which is less than the maximum response time for $RSSI$ reply. The proposed protocol is also aware of the mobility of the vehicle. The vehicle nodes may move towards the same direction or the opposite direction. Also, there is a possibility of moving to another channel due to PUs activity. Therefore, the location of the nodes may change and causes link failure or packet drop. In the proposed model, for avoiding link failure due to high mobility the neighboring node’s angle and speed are foretold. 

The link quality is measured in terms of delay and later defined as the transmission stability factor for the network. The source node selects the next neighboring node with a minimum delay which has better link quality. Here, every node is aware of traffic density as sharing the same channel by the node triggers congestion in the high-density area. Hence, a naive congestion avoidance mechanism is assumed to prevent interference problems between the vehicle nodes. Then, several internal or external attacks may degrade the performance of vehicular communication such as DoS, Black-hole, jamming, etc. So a reliability factor is implemented so that the neighboring nodes can trust received messages during the route discovery process which is shown in Algorithm 1. Initially, the Reliability Factor ($RF$) is 1 which indicates highly malicious and every query or suspect by the neighboring node this $RF$ value increments. When the $RF$ becomes infinity, the $NHDF$ becomes 0 by Equation 12 and is considered as malicious and discards from the path array. Therefore, the proposed routing protocol achieves secure and reliable data transmission. Then, A route maintenance mechanism is also implemented in the proposed routing protocol to ensure transmission without any link failure or packet drop and during multiple path discovery.

\begin{table}[ht]
\caption{Symbols used in the proposed routing protocol}
\label{tab:title}
\begin{tabular}{ll}
    \hline
      \textbf{Symbols} & \textbf{Description} \\
      
      \hline
      $VN_i$ & Any vehicular node in the network\\
      $VN_s$ & Source node or sender \\
      $VN_d$ & Destination node or receiver \\
    $RREQ$ & Route Request Message for data transmission \\
    $RREP$ & Route Reply Message for any $RREQ$\\
     $RF$ & Reliability factor of a vehicular node\\
     $RN$ & Number of Reports \\
      $Q_t$ & $50\%$ query or suspect by any vehicle node \\
      $L_s$ & Path size\\
      $route$ & 2D path array\\

    $SQN$ & Suspect or query from any vehicle node \\

      $Weight_{max}$ & Maximum values of $NHDF$ and $RF$ \\

      \hline
    \end{tabular}
\end{table}

\section{Proposed Routing Protocol}

This section highlights the development process of the proposed routing protocol. Hence, the section discusses on the routing metric, route discovery and selection mechanism along with route maintenance scheme. 

\subsection{Routing Metric}

The proposed routing protocol introduces routing metric $NHDF$. Here, total cumulative $NHDF$ $(N_p)$ is calculated regarding path $P$ which is expressed as follows,

\begin{equation}
      N_p = \sum_{i,j\in P_{th} s,d{} } N_{i,j}  
\end{equation}

Initially, the source node broadcasts the $RREQ$ message to its neighboring vehicular nodes. Then from the $RSSI$ value distance is measured to identify the current location of the neighboring nodes when the neighbor node receives the message which is shown below,

\begin{equation}
    d = 10^{(\frac{\kappa - 20log(\frac{4\pi l_0}{\upsilon})}{10\omega})}\ l_0
\end{equation}

where $\kappa$ is considered as path loss, $\omega$ is the signal loss exponent, $\upsilon$ is the wavelength of the received signal and $l_0$ is reference distance. 

Any vehicle upon receiving vehicle message stores a received timestamp \textit{$T_1$} along with the current co-ordinates. The node then waits for a random back-off period to send the reply. While replying, it sends sending time \textit{$T_2$} and current co-ordinates while sending. Here, the speed (\textit{s}) of the vehicle node during transmission can be estimated as follows,

\begin{equation}
s = \frac{\sqrt{(\alpha_r - \gamma_r)^2 + (\alpha_t - \gamma_t)^2 }}{(T_2 + \Delta ) - T_1}
\end{equation}

here, $\psi_r(\alpha_r, \gamma_r)$ and $\psi_s(\alpha_t, \gamma_t)$ are the co-ordinates of the vehicle node upon receiving and sending respectively where $\Delta$ is considered as transmission time.
Accordingly, if destination positions are $\varphi_r(\alpha_s, \gamma_s)$ and $\varphi_s(\alpha_q, \gamma_q)$ while receiving and sending packets respectively, the direction between neighbor and destination nodes can be calculated as follows,

\begin{equation}
    \theta = \cos^{-1}(\frac{\overrightarrow{\psi_r  \psi_s}\overrightarrow{\varphi_r \varphi_s}}{||\overrightarrow{\psi_r \psi_s}||\times ||\overrightarrow{\varphi_r \varphi_s}||})
\end{equation}
 
The displacement ($\tau_v$) between the two neighboring nodes during sending the data packet at time $T_2$ can be measured as follows,

\begin{equation}
    \tau_v = d\theta
\end{equation}

Therefore, during data transmitting between vehicular nodes the transmitting weight value ($\xi_T$) computed by every vehicle node is expressed as follows,

\begin{equation}
    \xi_T = \frac{\Phi_t}{\tau_v * \delta^L_P * s}
\end{equation}

here, $\Phi_t$ is considered as the transmission range of the vehicle node, $\tau_v$ is displacement between two neighboring nodes, $\delta^L_p$ is the cumulative sum of link delays in the route and s denotes the speed of the vehicle during transmission.

The proposed protocol supports three types of delays such as back-off delay, queuing delay and switching delay for identifying link failure between the intermediate or relay nodes. During transmission, traffic density plays an important role between the vehicular nodes within the transmission range. When data packet is transmitted through a highly dense area, it may be required to remain in the queue for a long period for forwarding the packet. Hence, if the number of neighboring vehicle nodes is determined as $V_i$, $RT_i$ is denoted as data rate of $N_i$ and \textit{S} is the size of the packet, the queuing delay $(\delta_L^N)$ of $N_i$ can be estimated as follows,

\begin{equation}
    \delta_i^N=\frac{SV_i}{RT_i}
\end{equation}

In the proposed approach for avoiding a collision, the vehicle nodes use random back-off time as multiple vehicle nodes may use the same channel. The back-off delay $(\delta_i^K)$ can be calculated through the following equation,

\begin{equation}
    \delta_i^K= \frac{1}{(1 - b_c)(1 - (1-b_c)^{V_i-1})}\ z 
\end{equation}

here, $b_c$ is identified as the probability of collision and $V_i$ is the neighboring vehicle node on a channel $CH_i$ and $z$ is the window size.

Accordingly, for the neighboring nodes discovery process, switching from one channel to another may be required for any intermediate node for forwarding the message to its next hop. Hence, a non-zero value is considered as the required time during this channel switching by a node. Switching delay $(\delta_{i,j}^M)$ is defined as the channel switching time and the corresponding position of the two channels in its channel-group dependent on it. Therefore, if a node $N_i$ is required to switch from channel p to channel q in its channel-group during forwarding the message to next-hop $N_j$, the estimating channel switching delay is defined as follows,

\begin{equation}
    \delta_{i,j}^M = a* |p - q |
\end{equation}

where $a$ is a positive real number and for a particular step size $a$ is considered as the tuning delay of two neighboring channels. For step size 10MHz, $a$ is considered to be 10ms \cite{Nafees_RARE}.

Thus, the delay of the link $(\delta_{i,j}^E)$ that connects $N_i$ and $N_j$ is calculated using Equation 7, 8 and 9 as below,

\begin{equation}
    \delta_{i,j}^E= \delta_{i,j}^M+ \delta_i^N+ \delta_i^K
\end{equation}

The proposed routing protocol also ensures secure data transmission by introducing RF factor. The reliability factor ($RF$) of the neighboring node can be estimated as follows,

\begin{equation}
    RF = e^{RN}
\end{equation}

here, $RN$ denotes the report number during every query or suspect by vehicular nodes during transmission.

Here, in the network $NHDF$ value, $N$ is calculated by each cognitive node and more importance is given on the $\xi_T$ value which accumulates robustness during transmission as the vehicular nodes select the maximum $\xi_T$ in the routing path where $C_n$ is the number of common idle channels, link delay $(\delta_{i,j}^E)$ and $RF$ is the reliability factor .

\begin{equation}
     N _{i,j}= \frac{(\frac{\xi_T}{\delta_{i,j}^E})^{C_n}}{RF}
\end{equation}

\subsection{Route Discovery Process}

Route discovery process in the proposed protocol is presented in Algorithm 1 where a source node initiates the discovery process by broadcasting the $RREQ$ to the neighbors. Then the neighbor nodes check if the intermediate node is reliable or not. The neighboring nodes then perform the query about the nodes. Upon receiving each suspect or query the RF factor increases. If the $SQN$ is more than 50\% valid for a particular vehicle node, the node is then detected as malicious and every neighboring node involved in the query process sets the node’s value infinity. Then during the data transmission process, every vehicle node can identify this malicious node by the infinity value and discards the node during route selection. During $RREQ$ if the destination node exists in the neighboring list, the destination node calculates $NHDF$ using Equation 12 associated with distance, speed ratio, angle and number of free channel. After that, the destination node sends a route reply message with distance, angle, speed ratio, number of free channel and $RF$ value to the previously generated $RREQ$ neighboring node. After receiving the $RREP$ message by the neighboring node, it then again measures $NHDF$, $RF$ values, free common channels and after adding the path with the link the neighbor then sends $RREP$ message to the previously generated $RREQ$ neighboring node. The process continues til the neighbor node receives $RREP$ message and all the possible routing paths are discovered. The source node calculates $NHDF$ and $RF$ values and stores routing path with $NHDF$ and $RF$ in the route array.

\subsection{Route Selection and Maintenance Process}
The route selection process is presented in Algorithm 2 in which the source node selects the routing path for data transmission from the source node to the destination node. When the source node discovers multiple routes, then it selects the path associated with the least $NHDF$ and $RF$ value. Then the source node transmits packet to the destination using this selected route. However, if during the route discovery process, only one routing path is discovered, the source node selects the routing path to the destination.

For route maintenance of the proposed routing protocol, link failure is taken into consideration. When a link failure occurs, the proposed protocol discovers a new path for route selection by sending a message of route error by link failure predecessor node. Then this path link is deleted by the source node from the route array. Afterward, this new routing path is chosen to transmit packet to the destination. 

\begin{figure}[!t]
\removelatexerror
 \label{algo1}
  \begin{algorithm}[H]
  \caption{Route Discovery Algorithm}

  \textbf{START} \\
  $RF = 1$ \\
  $VN_s$ broadcasts $RREQ$ \\
    \While{(RREQ)}
    {
      $VN_i$ do $SQN$ for non-reliability status \\
        \While{($SQN$)}{
            SET $RF$ = $RF$ +1 \\
            \uIf{($RF <$  $Q_t$) }{
                $RF$ remains same
            }
            \uElse{
                SET $RF$ = $INFINITY$ \\

            }   \textbf{end}

        \uIf{($VN_d \in VN_i$)}{
        $VN_i$ calculates $NHDF$ using Eq. 12 \\
         $VN_i$ sends RREP with $NHDF$ and $RF$ with path link\\
            
            \uIf{($VN_i$ $!= VN_s$)}{
             $VN_i$ calculates $NHDF$ using Eq. 12 \\
             $VN_i$ sends RREP with $NHDF$ and $RF$ with path link\\

        } \uElse {
            \textbf{Break} \\
        }
        \textbf{end}
        
             \uIf{($VN_i$ $= = VN_s$)}{
              $VN_i$ calculates $NHDF$ using Eq. 12 \\
              $VN_s$ stores path with $NHDF$ and $RF$ in routing path\\
        } \uElse{
                    \textbf{Break} \\

        }
        \textbf{end}
        
        }\uElse{
            $VN_i$ broadcasts $RREQ$ \\
        }
        \textbf{end}\\
        
        }
        \textbf{end}\\

      } 
    \textbf{end}\\

        \uIf{($L_s >$ 1)}{
        Path\_Selection = Path\_Selection\_Algorithm \\
    $VN_s$ sends message to $VN_d$ using Path\_Selection \\
        
        }
        \uElse{$VN_s$ sends message to $VN_d$ using path\\
        }         \textbf{end}

\textbf{END}
\end{algorithm}
\end{figure}

\begin{figure}[!t]
 \removelatexerror
  \begin{algorithm}[H]
  \caption{Route Selection Algorithm}
  \textbf{START} \\
  $Weight_{max}$ = $route [1,1]$\\
  $flag$ = 1\\
  \While{($n=2:L_s >$ $Weight_{max}$)}{
        
    \uIf{($route[n,1] > Weight_max$)} {
            $Weight_{max}=$ $Route[n,1]$\\
            flag = n\\
        }
    \uElse{
            Break \\
    }
        \textbf{end}
  }
    \textbf{end}\\
\textbf{return} $route[flag,2]$ \\
\textbf{END}

\end{algorithm}
\end{figure}

\section{Simulation Results}

The performance of the proposed routing protocol has been evaluated in the simulation environment using NS2, a discrete-event simulator. During the simulation,  every observation has a run-time of 150 seconds and the average results from several observations have been considered for the comparative study. Hence, the performances of the proposed routing protocol have been compared with that of three other recently developed routing protocols namely, PGRP \cite{Karimi}, SDGR \cite{Ji}, and MPBRP \cite{Ye} in this section. 

Moreover, packet delivery ratio, throughput, and average end-to-end delay are considered to be the metrics for performance assessments of these protocols. Here, the packet delivery ratio is defined as the total number of delivered data packets at the destination over the total packets sent during the transmission from a source node. Besides, throughput is defined to be the number of successfully delivered packets per unit time at the destination. On the other hand, the end-to-end delay is measured as the time required for carrying data packets from the source node to the destination node. 

\begin{table}[ht]
\caption{Simulation Environments}
\label{tab:simu}
\begin{center}

\begin{tabular}{ll}
    \hline
    \textbf{Parameters} & \textbf{Value} \\
      
     \hline
      Number of vehicular nodes & 120,140,160,180,200\\
      Simulation area & 4000$m^2$ \\
      Number of Channels & 100 \\
      Run time & 150 s \\
      Packet size & 512 bytes \\
      Data transmission rate & 2 Mbps \\
      Traffic type & CBR \\
      Queue type & Drop-tail \\
      Transmission range & 500 m \\
      Maximum vehicular speed & 2 $m/s$ \\
      Propagation model & Two-Ray Ground \\
      Mac Layer Protocol & 802.11p \\
      Antenna type & omni-directional \\
      \hline
    \end{tabular}
  \end{center}
\end{table}

\begin{figure*}[t]
\centering
\mbox{\subfloat[Comparison in terms of Packet Delivery Ratio]{\label{pdr}\includegraphics[width=0.33\linewidth]{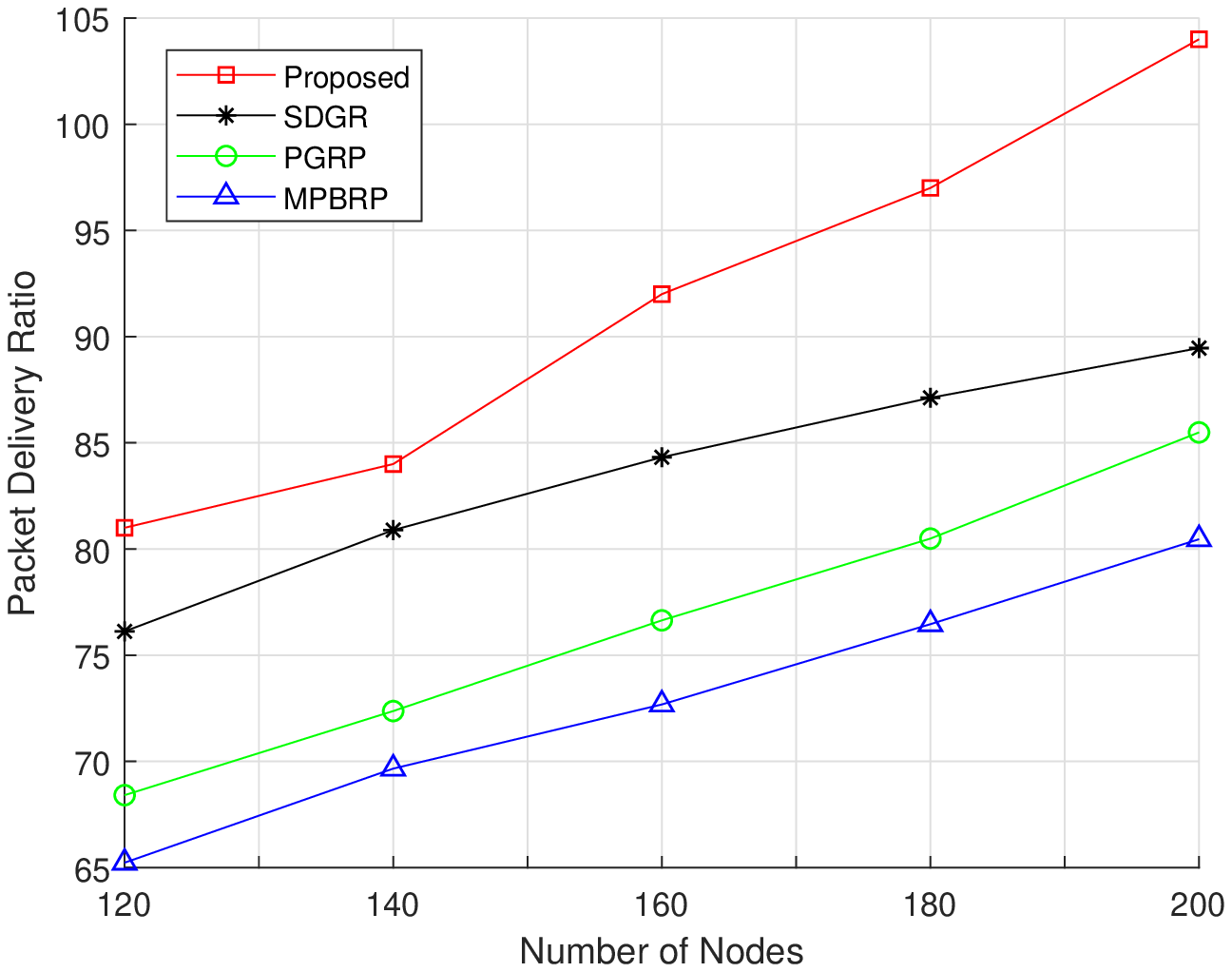}}}\hfill 
\mbox{\subfloat[Comparison in terms of End-to-End Delay]{\label{delay}\includegraphics[width=0.33\linewidth]{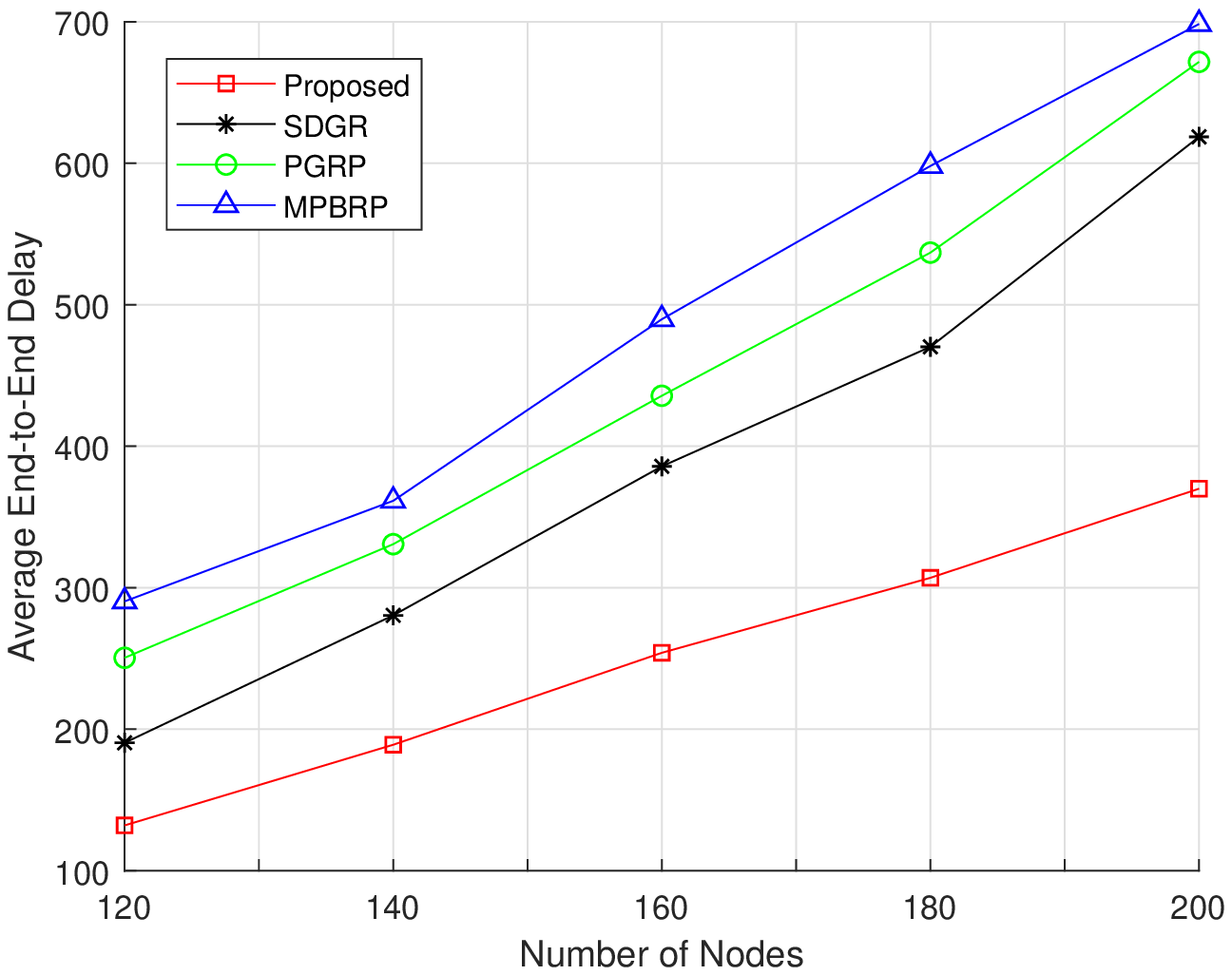}}}
\mbox{\subfloat[Comparison in terms of Throughput]{\label{throughput}\includegraphics[width=.33\linewidth]{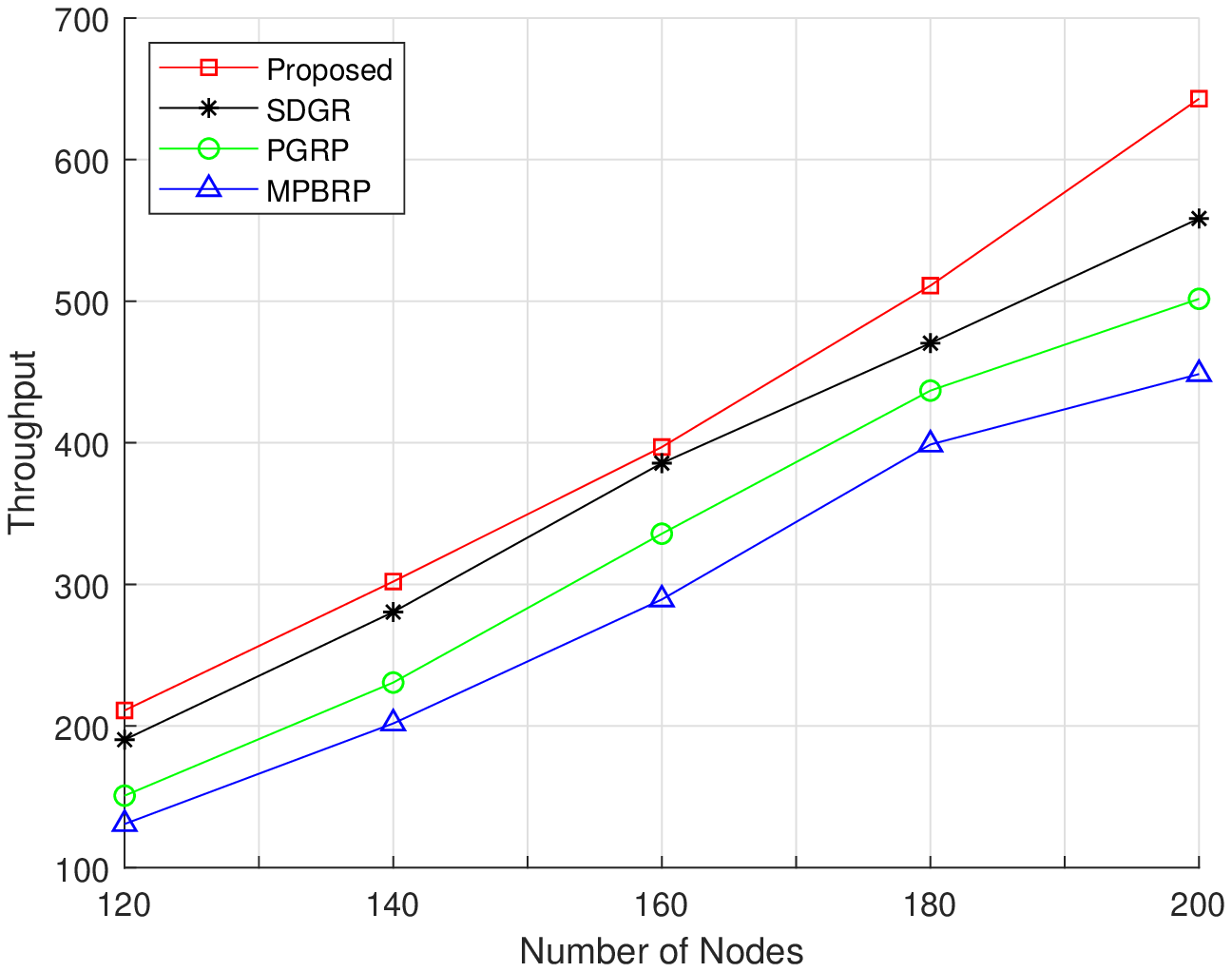}}}
\caption{Performance comparisons of the proposed routing protocol with other protocols.}
\label{simu}
\end{figure*}

Here, Fig. 1(a) depicts the performances based on the packet delivery ratio for the proposed routing protocol along with the other three protocols. It has been portrayed in Fig. 1(a) that traffic density has a significant impact on the performance of all the protocols in terms of packet delivery ratio. It is also observed that the packet delivery ratio in every routing protocol increases as the traffic density increases. This is because, with the increasing number of nodes, the impact of the vehicles' mobility decreases for all the protocols. Thus, the rate of successful transmission is increased while the overall packet drops are reduced. Fig. 1(a) also indicates that the proposed routing protocol performs better from the viewpoint of the packet delivery ratio than that of PGRP, SDGR, and MPBRP. This is because; in CR-VANET, the changing availability of the channels is also a performance-limiting factor and a precondition for the stable route. Since the number of shared channels is also considered during the path selection process in the proposed protocol, links in the proposed routing protocol are more stable and achieve higher successful transmissions. However, the other routing protocols suffer from a higher packet dropping rates due to a lack of stability in the connection between vehicular nodes under varying availability of the channels.
Fig. 1(b) shows that end-to-end delay increases with the increasing number of nodes in all four routing protocols. This is because, in a dense network, a higher number of intermediate nodes are involved in the communication process. This results in increased individual data processing sessions. Moreover, the proposed routing protocol performs better for end-to-end delay compared with the other protocols. This is because,  three different types of delay are considered in the route selection process in the proposed routing protocol, where the protocol attempts to select a routing path with the minimum delay. 
Fig. 1(c) demonstrates that throughput in all the routing protocols is increased for the increased traffic density. This is because nodes' mobility has a lesser impact in the dense network resulting in the connections among vehicles to be more stable. Moreover, the proposed routing protocol performs better compared to PGRP, SDGR, and MPBRP in terms of network throughput. This is because the proposed protocol selects the optimal route with higher link stability for varying availability of the spectrum. Moreover, the proposed protocol also ensures faster data delivery by considering three different types of delays in the routing metrics. On the other hand, selected routes in PGRP, SDGR, and MPBRP are less stable in varying availability of the spectrum. Hence, selected routes in these protocols eventually result in higher packet drops and link failures.

\section{Conclusion and Future Works}

In this article, a spectrum aware mobility pattern based reactive routing protocol for CR-VANET is presented. The proposed protocol considers the double-folded dynamic behavior of the network which are autonomous movements of the vehicles and changing spectrum availability. In the proposed multi-hop routing protocol, the weight for an edge is measured based on a parameter named $NHDF$ (Next-hop Determination Factor), where the $NHDF$ implicitly considers mobility pattern, number of shared channels, link quality, and node's reliability. From the simulation results, it is observed that the proposed routing protocol performs better than other recently introduced protocols. This study will lead to the further development of the routing protocol to be energy efficient.

\end{document}